\documentclass[aps,pra,twocolumn,superscriptaddress]{revtex4}

\usepackage{graphicx}

\begin{document}

\title{The effect of dissipation on quantum transmission resonance}

\author{Kohkichi Konno}
\affiliation{Department of Applied Physics, Hokkaido University,
                 Sapporo 060-8628, Japan.}
\author{Munehiro Nishida}
\affiliation{Graduate School of Advanced Sciences of Matter,
              Hiroshima University, Higashi-Hiroshima  739-8530, Japan.}
\author{Satoshi Tanda}
\email[]{tanda@eng.hokudai.ac.jp}
\affiliation{Department of Applied Physics, Hokkaido University,
                 Sapporo 060-8628, Japan.}
\author{Noriyuki Hatakenaka}
\affiliation{Graduate School of Integrated Arts and Sciences,
              Hiroshima University, Higashi-Hiroshima  739-8521, Japan.}

\date{\today}

\begin{abstract}
Quantum transmissions of a free particle
passing through a rectangular potential barrier with dissipation
are studied  using a path decomposition technique.
Dissipative processes strongly suppress the transmission probability at
resonance just above the barrier
resulting in an unexpected reduction of the mean traversal time through the
potential barrier. 
\end{abstract}


\maketitle

\section{Introduction}

Quantum mechanics successfully describes 
physical processes at the microscopic
scale and sometimes exhibits unique counter-intuitive phenomena.
One such example is quantum-mechanical tunneling 
that cannot be explained
in terms of classical mechanics. In search of
an applicability for quantum mechanics 
on a macroscopic scale \cite{leggett1}, 
the effect of dissipation on quantum tunneling 
has been studied in respect to 
systems such as superconducting devices \cite{cl1,cl2,voss} because  
dissipation is inherent and inevitable in macroscopic scale. 
The observation of quantum-tunneling-rate reduction 
was a first important result 
to be associated with quantum mechanics on a macroscopic scale.

\begin{figure}
     \centering
     \includegraphics[width=1.0\linewidth]{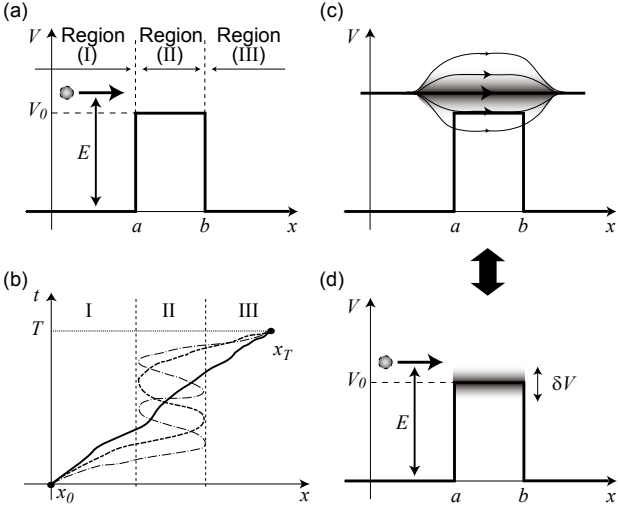}
     \caption{\label{fig:potential}
       Schematic diagrams of (a) three decomposed regions for
       a rectangular potential barrier,
       (b) typical possible paths, (c) uncertainty of energy,
       and (d) uncertainty of potential height, where
       the diagram (c) is equivalent to (d).}
\end{figure}

Quantum tunneling, however, is not the only intrinsic feature
of quantum mechanics. Another unconventional example 
in a classical sense is {\it quantum reflection}.
Suppose that there is a particle incident to a potential barrier, which
has a height slightly lower than the particle's energy,
as shown in Fig.~\ref{fig:potential}(a).
According to classical mechanics,
the particle overcomes the barrier and is never reflected back.
In quantum mechanics, there is the possibility that the particle cannot
go over the barrier. 
This leads to a remarkable phenomenon called {\it a transmission resonance}. 
That is due to multiple quantum reflections between potential boundaries.
In other words, the resonance occurs due to {\it interference} 
associated with the back-and-forth motions of the particle.  
Quantum tunneling only shows its quantum-mechanical 
feature at the moment of the tunneling event, 
while the interference requires for a certain 
definite period to retain quantum coherence 
based on the quantum-mechanical superposition of states. 
Therefore, 
quantum interference appears a more convincing quantum-mechanical effect 
than does quantum tunneling. 
Indeed, the effect of dissipation on the quantum-mechanical
superposition of macroscopically distinguishable states, 
so-called Schr\"{o}dinger's cat in a fundamental 
problem of quantum mechanics, 
has been studied along these lines in respect to 
superconducting nanodevices 
\cite{friedman,vanderwal} for example. 
Recently, this cat state has been utilized as 
a building block for a quantum computer in quantum information science.
Furthermore, direct observations of matter wave interference
revealed that inevitable sources of dissipation for large molecules,
e.g. collisions to external molecules \cite{hornberger} or internal
vibrations resulting in thermal photon radiations \cite{hackermuller},
play a key role for the quantum-to-classical transition of ``free''
particle.
Even the gravitational waves were suggested to become a source of
decoherence of matter waves \cite{lamine}.
Thus, macroscopic transmission resonance originated from quantum
interference of macroscopic object also provides 
an alternative platform to that of quantum tunneling 
for testing the validity of quantum mechanics on a macroscopic scale.

The transmission of a particle through a rectangular potential 
barrier {\em in the absence of dissipation} has been well studied 
within the framework of Schr\"odinger's wave 
mechanics at the textbook level.  
A simple extension for incorporating dissipation 
in that framework was made by Cai et al.\cite{cai}
to deal with the problem that an electron 
propagates above a quantum well 
with dissipation due to the electron--optical-phonon interaction. 
They succeeded in revealing the electron-capture process
in a quantum well that involves a loss of electron energy
via phonon emission.
However, in order to investigate the problem addressed here,
we need an alternative approach in which the influence of the particle
motion on the environment is also included.
In particular, our interest is directed to the 
transmission resonance formed just above the potential barrier where 
the Wentzel-Kramers-Brillouin (WKB) method adopted in previous 
studies \cite{bb,bp1} breaks down.

In this Letter, we employ a path decomposition expansion method 
\cite{path-d,tt-path,sz,fertig} based on the path-integral approach, 
and develop it to incorporate dissipative processes. 
Then we discuss the effect of dissipation 
on quantum transmission resonance. 
Since the resonance may be attributed to the back-and-forth 
motions of a particle 
between the potential boundaries, 
as pointed out by Bohm in his seminal book \cite{bohm}, 
the particle will stay in region II for a longer time at resonance.  
Thus the resonance could be characterized by the time 
spent in the potential barrier region, i.e., the traversal time. 
Therefore, we investigate dissipative quantum transmission resonance 
in terms of traversal time on the basis of Bohm's interpretation.

This Letter is organized as follows. 
In Section \ref{sec:tp}, we formulate  
the transmission probability through a 
rectangular potential barrier with dissipation by using a 
path decomposition technique. We also perform 
numerical calculations of the transmission probabilities. 
Then we will find the non-uniform reduction of transmission probability 
even though we assume an energy-independent damping processes. 
In Section \ref{sec:tt}, 
We introduce the traversal time under dissipation 
to explain the unexpected reduction 
on the basis of Bohm's interpretation. 
We also provide further evidence for our 
interpretation by using traversal time distribution, which
results in an unexpected shortened mean traversal time. 
In Section \ref{sec:summary}, 
we provide a summary and propose a possible experimental setup.

\begin{figure}
     \centering
     \includegraphics[width=1.0\linewidth]{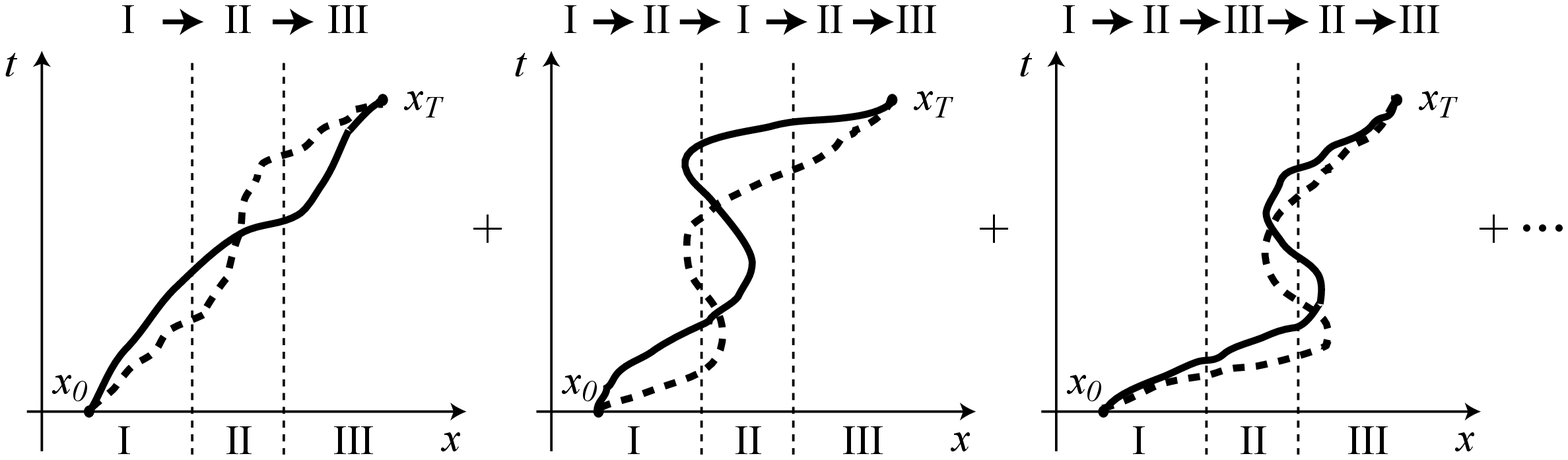}
     \caption{\label{fig:path-d}
       Illustration of the path decomposition technique.
       Several groups of summations of all possible paths
       are shown (see also text).
       Each group is classified according to which regions
       the paths have passed over in terms of temporal development.
       Two examples of the possible paths
       are shown by solid and dashed curves in each figure.}
\end{figure}

\section{Transmission Probability in the Presence of Dissipation}
\label{sec:tp}

\subsection{Analytical description of transmission probability}
We briefly review a path decomposition expansion developed by Auerbach
and Kivelson \cite{path-d}. 
This enables us to deal with quantum transmission 
in terms of the path integral approach. 
In the path decomposition technique, the summation of
possible paths is decomposed into certain groups as shown in
Fig.~\ref{fig:path-d}.
The first group is composed of all possible paths that
pass straight over regions I, II, and III in that order.
The next group includes all paths that go from region I
to region II, then return to region I once,
and go from region II to III. In the same way,
one can consider all other groups. 
Thus, the total summation of paths is given by 
an infinite series of groups as shown in Fig. 2. 
Each group is completely expressed by
the propagators 
$K^{({\rm I})}(x',x)$, $K^{({\rm II})}(x',x)$ 
and $K^{({\rm III})}(x',x)$
defined in the restricted regions I, II and III, respectively.
Since all possible paths are taken for summation,
no approximation is used in this technique,
the point of which prevails against the WKB method.
The propagator $K(x_{T},x_{0};T)$
from $x_{0}(<a)$ at $t=0$ to $x_{T}(>b)$ at $t=T$
is then decomposed as \cite{path-d}
\begin{eqnarray}
\label{eq:path-d}
  \lefteqn{iK(x_{T},x_{0};T)} \nonumber \\
   & = & \int_{0}^{T} dt_{1} \int_{0}^{T-t_{1}} dt_{2} \:
      i K^{({\rm I})} (a,x_{0};t_{1}) 
      \Sigma_{x}^{a} \left( i K^{({\rm II})} (b,x;t_{2}) \right)
      \nonumber \\
   &&  \qquad \times
      \Sigma_{x}^{b} \left( i K^{({\rm III})} (x_{T},x;T-t_{1}-t_{2})
      \right) \nonumber \\
   &&  \quad + \int_{0}^{T} dt_{1} \int_{0}^{T-t_{1}} dt_{2}
      \int_{0}^{T-t_{1}-t_{2}} dt_{3} 
      \int_{0}^{T-t_{1}-t_{2}-t_{3}} dt_{4} \nonumber \\
   && \qquad \times i K^{({\rm I})} (a,x_{0};t_{1}) 
      \Sigma_{x}^{a} \left( i K^{({\rm II})} (a,x;t_{2}) \right)
      \nonumber \\
    && \qquad \times 
      \Sigma_{x}^{a} \left( i K^{({\rm I})} (a,x;t_{3}) \right)
      \Sigma_{x}^{a} \left( i K^{({\rm II})} (b,x;t_{4}) \right)
      \nonumber \\
    && \qquad \times 
      \Sigma_{x}^{b} \left( i K^{({\rm III})} 
      (x_{T},x;T-t_{1}-t_{2}-t_{3}-t_{4}) \right) \nonumber \\
    && \quad 
      + \cdots ,
\end{eqnarray}
where 
$\Sigma_{x}^{a}$ denotes a derivative operator defined by
\begin{equation}
\label{eq:sigma-o}
 \Sigma_{x}^{a} \left( i K^{({\rm II})} (b,x;t) \right)
  \equiv \left. \epsilon_{ab} \frac{\hbar}{2m} 
   \frac{\partial}{\partial x} 
   \left( i K^{({\rm II})} (b,x;t) \right) \right|_{x=a} .
\end{equation}
Here, $m$ is the particle mass, and
\begin{equation}
 \epsilon_{ab} \equiv \left\{ 
 \begin{array}{cc}
  1 & (a<b)  \\ -1 & (a>b) 
 \end{array} \right. 
\end{equation}
(see Ref.~\cite{path-d} for details).

In the absence of dissipation, 
the propagator $K^{(\rm II)}$ in region II
is expressed as \cite{schulman}
\begin{eqnarray}
\label{eq:KII}
 \lefteqn{K^{({\rm II})}(x',x ;t)}\nonumber \\
 & = & \sqrt{\frac{m}{2\pi i \hbar t}} 
     \sum_{n=-\infty}^{\infty} \left\{ 
     \exp \left[ \frac{im (2nd +x'-x)^2}{2\hbar t}\right]
     \right. \nonumber \\
 && \quad \left. 
   - \exp \left[ \frac{im ((2n+1) d - (x'-b)-(x-a))^2}{2\hbar t}\right]
    \right\} ,
\end{eqnarray}
where $d$ is defined by $d=b-a$, and 
$n$ characterizes different classical paths in each term.
The first term is composed of all paths with
even numbers of reflections at the boundaries of $x=a$ and $x=b$,
and the second term corresponds to the odd numbers of reflections.
Paths that start to move in a negative direction 
in relation to the initial position are characterized by negative $n$.
Equation~(\ref{eq:KII}) indicates that the propagator
in region II is essentially expressed by the summation of 
a free particle's propagator; 
\begin{eqnarray}
\label{eq:prop-k0}
 K_{0} (x',t;x,0) 
 & = & \sqrt{\frac{m}{2\pi i \hbar t}}
   \exp \left[ \frac{im(x'-x)^2}{2\hbar t} \right] . 
\end{eqnarray}

Now let us consider the propagator $K_{\rm D}^{(\rm II)}$ 
in the presence of dissipation. 
Dissipation in quantum mechanics has long been discussed since 
it cannot be included as a form of analytical mechanics. 
It remains unresolved. 
However, several aspects such as dissipative quantum tunneling 
and quantum Brownian motion 
have been presented. 
Here we employ a phenomenological model successfully 
introduced by Caldeira and Leggett \cite{cl}, 
to describe dissipation in a study of quantum 
Brownian motion of a particle in harmonic potential. 
They modeled an environment as a set of 
a huge number of harmonic oscillators that produces a classical 
equation of motion with dissipation.  
We apply their model to a free particle coupled to the environment.

According to their model, 
the effect of the propagator $K_{\rm D}^{(\rm II)}$ is included 
in the expression for the time evolution of the system of interest 
coupled to environment described by 
\begin{equation}
 \rho (x,y,t) = \int dx_{\rm i}dy_{\rm i}
 J(x,y,t;x_{\rm i},y_{\rm i},0) \rho (x_{\rm i},y_{\rm i},0) ,
\end{equation}
where $J(x,y,t;x',y',0)$ is the propagator for the density matrix 
$\rho (x,y,t) = \left< x | \psi (t) \right> \left< \psi (t) | y\right>$ 
of the free particle, and the autocorrelation of stochastic
force $F_{\rm cl}(\tau )$ 
\begin{equation}
\label{eq:f-crr2}
 \left< F_{\rm cl}(\tau ) F_{\rm cl}(s) \right> = 2 \eta k_{\rm B}T'
  \delta \left( \tau -s \right) 
\end{equation}
is imposed on it. 
Indeed, in the absence of dissipation, the above expression 
for a free particle includes the free particle's 
propagator $K_{0}$ as
\begin{equation}
 \rho (x,y,t) = \int dx_{\rm i}dy_{\rm i}
 K_{0}^{\ast}(y,t;y_{\rm i},0) K_{0}(x,t;x_{\rm i},0)  
 \rho (x_{\rm i},y_{\rm i},0) .
\end{equation}
In particular, when the initial state is given by 
$\left< x | \psi (0) \right> = \delta (x-x_{0})$, 
we simply have
\begin{equation}
 \rho (x,x,t) = 
 K_{0}^{\ast}(x,t;x_{0},0) K_{0}(x,t;x_{0},0)  
 = \frac{m}{2\pi i \hbar t}.
\end{equation}

In the case of a free particle coupled to a set of harmonic oscillators,
with $\left< x | \psi (0) \right> = \delta (x-x_{0})$, 
the density matrix is given by 
\begin{equation}
\label{eq:rho-xx}
 \rho \left( x , x ,t \right)
  = \frac{m}{2\pi \hbar t} f(t), 
\end{equation}
with
\begin{equation}
 f(t) \equiv  \frac{\sigma t e^{\gamma t}}
     {\sinh \sigma t} ,
\end{equation}
where $\gamma$ is a relaxation rate, and 
$\sigma$ is defined by 
$\sigma \equiv \sqrt{\gamma^2 + (4\gamma \Omega / \pi )^2}$.
Here $\Omega$ is the cutoff frequency for the frequency 
distribution of harmonic oscillators.
The function $f(t)$ can be obtained 
by integrating the degrees of freedom of the environment
and, therefore, includes the effect of 
the associated harmonic oscillators.
Since the cutoff frequency $\Omega$ 
is sufficiently large compared with $\gamma$,
the function $f( t )$ becomes a monotonically decreasing
function, which decays exponentially, i.e., 
$f( t ) \propto e^{- ( \sigma - \gamma )t}$.
It should also be noted that $f(0)=1$ and $f(t)>0$ for $t>0$.
In the limit of $\gamma \rightarrow 0$, since  
$f(t) \rightarrow 1$, we retrieve the result for
the non-dissipative case
$| K_{0} (x,t;x_{0},0) |^2 = m/2\pi \hbar t $,
where $K_{0}$ is the free particle's propagator 
shown in Eq.~(\ref{eq:prop-k0}).
Therefore, we can consider 
the propagator for the dissipative case 
to be effectively expressed as 
\begin{eqnarray}
\label{eq:K_D}
 K_{\rm D} (x,t;x_{0},0) 
 & = & \sqrt{f(t)} K_{0} (x,t;x_{0},0) ,
\end{eqnarray}
which gives the same result as Eq.~(\ref{eq:rho-xx}).

The propagator in region II is obtained 
by summing up the free particle's propagators along possible 
classical paths. Thus, we obtain the effective propagator 
in region II with dissipation as 
\begin{eqnarray}
\label{eq:KII_D}
 K_{\rm D}^{({\rm II})}(x',x ;t)
  = \sqrt{f(t)} K^{({\rm II})}(x',x ;t) .
\end{eqnarray}
The full propagator, taking 
account of the dissipative effect, is given by 
\begin{eqnarray}
\label{eq:path-d-d}
  \lefteqn{iK_{\rm D}(x_{T},x_{0};T)} \nonumber \\
   & = & \int_{0}^{T} dt_{1} \int_{0}^{T-t_{1}} dt_{2} \:
      \sqrt{f(t_{2})} i K^{({\rm I})} (a,x_{0};t_{1}) 
      \nonumber \\
   &&  \qquad \times
      \Sigma_{x}^{a} \left( i K^{({\rm II})} (b,x;t_{2}) \right)
      \nonumber \\
   &&  \qquad \times
      \Sigma_{x}^{b} \left( i K^{({\rm III})} (x_{T},x;T-t_{1}-t_{2})
      \right) \nonumber \\
   &&  \quad + \int_{0}^{T} dt_{1} \int_{0}^{T-t_{1}} dt_{2}
      \int_{0}^{T-t_{1}-t_{2}} dt_{3} 
      \int_{0}^{T-t_{1}-t_{2}-t_{3}} dt_{4} \nonumber \\
   && \qquad \times \sqrt{f(t_{2}+t_{4})}
      i K^{({\rm I})} (a,x_{0};t_{1}) 
      \Sigma_{x}^{a} \left( i K^{({\rm II})} (a,x;t_{2}) \right)
      \nonumber \\
    && \qquad \times 
      \Sigma_{x}^{a} \left( i K^{({\rm I})} (a,x;t_{3}) \right)
      \Sigma_{x}^{a} \left( i K^{({\rm II})} (b,x;t_{4}) \right)
      \nonumber \\
    && \qquad \times 
      \Sigma_{x}^{b} \left( i K^{({\rm III})} 
      (x_{T},x;T-t_{1}-t_{2}-t_{3}-t_{4}) \right) \nonumber \\
    && \quad 
      + \cdots ,
\end{eqnarray}
where since $f(t)$ behaves as an exponentially decaying function,
we have used the approximation 
$f(t) f(t') \approx f(t + t')$, whose iterative use leads to 
$\prod_{i} f(t_{i}) \approx f(\sum_{i}t_{i})$.

To convert Eq.~(\ref{eq:path-d-d})
into an energy representation, we perform a
Fourier transform of the propagator as
\begin{equation}
\label{eq:define-green-f}
 G_{\rm D} (x_{T} , x_{0} ; E)
 \equiv i \int_{0}^{\infty} dT K_{\rm D} (x_{T},x_{0} ; T)
 e^{iET/ \hbar} .
\end{equation}
Utilizing the expression
\begin{eqnarray}
  \lefteqn{\sqrt{f \left( \Sigma_{i} t_{i} \right)}}
   \nonumber \\ 
   & = & \int_{0}^{\infty} \!\! d\tau \sqrt{f(\tau)} \: \delta
        \left( \tau - \Sigma_{i} t_{i} \right) \nonumber \\
   & = & \int_{0}^{\infty} \!\! d\tau \sqrt{f(\tau)} \frac{1}{2\pi}
        \int_{-\infty}^{\infty} d\omega
        e^{-i \omega \left( \tau - \Sigma_{i} t_{i}
        \right)} , 
\end{eqnarray}
we finally obtain the Green function 
including the dissipative effect as
\begin{equation}
\label{eq:gd}
 G_{\rm D} \left( x_{T} , x_{0} ; E \right)
     = w_{\rm D} \left( E , V_{0} \right)
       G_{0} \left( x_{T} , x_{0} ; E \right),
\end{equation}
where $G_{0} (x_{T},x_{0})$ is the Green function
when the barrier is absent, and the transmission amplitude
$w_{\rm D} \left( E , V_{0} \right)$ is given by 
\begin{eqnarray}
\label{eq:tr-amp}
 \lefteqn{w_{\rm D} \left( E , V_{0} \right)} \nonumber \\
 & = & \int_{0}^{\infty} d\tau \sqrt{f(\tau)} \int_{-\infty}^{\infty}
       \frac{d\omega}{2\pi} e^{-i\omega \tau}
       w \left( E , V_{0} - \hbar \omega \right) .
\end{eqnarray}
Here, $w(E,V_{0})$ is the transmission amplitude 
in the absence of dissipation
\begin{eqnarray}
\label{eq-w0}
 w(E,V_{0})
 = - \frac{2i k \kappa e^{-ikd}}
 {\left( k^2 + \kappa^2 \right) \sin \kappa d 
  + 2ik\kappa \cos \kappa d} \quad ,
\end{eqnarray}
where
\begin{equation}
 k \equiv \frac{\sqrt{2mE}}{\hbar}, \quad 
 \kappa \equiv \frac{\sqrt{2m(E-V_{0})}}{\hbar} .
\end{equation}
The paths in an energy representation are 
introduced by the potential deviations 
in terms of an energy quantum $\hbar \omega$ as shown in Fig. 1 (c) and (d). 
Therefore, the transmission probability including 
the effect of dissipation is given by
$| w_{\rm D} \left( E , V_{0} \right) |^2$.
Equations (\ref{eq:gd}) and (\ref{eq:tr-amp}) are 
our main result.

\subsection{Numerical estimates of transmission probabilities}

\begin{figure}
     \centering
     \includegraphics[width=0.7\linewidth]{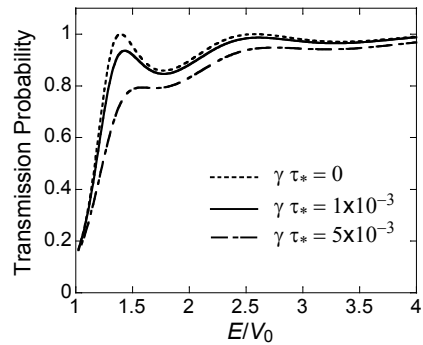}
     \caption{\label{fig:trans-prob}
        Transmission probabilities as a function of $E/V_{0}$.
        $\gamma \tau_{\ast} = 5 \times 10^{-3}$,
        $1 \times 10^{-3}$ and $0$ are plotted, where
        $\tau_{\ast} \equiv \sqrt{md^2 / 2V_{0}}$.
        In the calculations, we adopted $d/\lambda_{0}=5$ and
        $\Omega \tau_{\ast} = 100$.}
\end{figure}
In our numerical calculations, 
we restrict ourselves to a constant $\gamma$ damping for any $E$ 
even though the relaxation rate $\gamma$ may depend
on the incident energy $E$ of the particle in general. 
Figure \ref{fig:trans-prob} shows the transmission probabilities
for different $\gamma$ values.
Two dimensionless parameters characterize the transmission probability, 
i.e., the particle energy normalized by the potential height,
$E/V_{0}$, and the potential width normalized by the
typical scale of length, $d/\lambda_{0}$,
where $\lambda_{0} \equiv \hbar / \sqrt{2mV_{0}}$.
The dashed curve shows the transmission probabilities for 
a non-dissipative case.
The oscillatory structure is caused by transmission resonance, 
which arises as a result of quantum reflection 
at the barrier edges. 
 From  Eq.~(\ref{eq-w0}), 
perfect transmission occurs whenever the barrier 
contains an integer number of a half wavelength, i.e.,
$\kappa d = n\pi \; (n=1,2,\cdots)$.

In the presence of dissipation, the transmission 
probabilities are greatly suppressed.
The blurred oscillatory structure implies that the suppression 
is not uniform with respect to energy, even for
the energy-independent relaxation rate $\gamma$.
Indeed, significant suppression occurs 
around the resonance conditions.

\section{Traversal Time through Barrier}
\label{sec:tt}

\subsection{Traversal time and transmission probability}

In a classical description transmission resonance 
can be attributed to the back-and-forth 
motion of a particle between the edges of the potential barrier. 
This will lead to a longer stay in region II, 
equivalent to a longer traversal time.
Hence we can investigate the transmission probabilities 
in the presence of dissipation 
in terms of the traversal time through region II.

In terms of the path integral formalism \cite{tt-path,sz,fertig}, 
the traversal time has a distribution, 
because all possible paths are considered 
and a certain weight is assigned to each path in the path integral.
According to Fertig\cite{fertig}, 
the probability amplitude of a particle
spending time $\tau$ in region II is defined by
\begin{eqnarray}
\label{eq:t-time-amp}
  F(\tau ) & \equiv &
       \frac{\sum_{{\rm path}:C} e^{iS(C)/ \hbar}
         \delta (\tau - \tau_{\rm II}(C))}
       {\sum_{{\rm path}:C} e^{iS(C)/ \hbar}} ,
\end{eqnarray}
where $C$ denotes a path, $S$ denotes the action, 
and $\tau_{\rm II}(C)$ is the traversal time through 
region II along path $C$.
The denominator in Eq.~(\ref{eq:t-time-amp})
is by definition equivalent to the propagator 
$K(x_{T},x_{0};T)$.
The delta function in the numerator 
extracts the paths with traversal time $\tau$.
In the following discussion, we assume 
a constant energy state
within the limits of $x_{0} \rightarrow -\infty$ 
and $x_{T} \rightarrow \infty$.

First, we investigate the relationship between the transmission 
resonance and the traversal time in the absence of dissipation.
The mean traversal time $\left< \tau \right>$ 
is defined by
\begin{equation}
 \left< \tau \right> \equiv \int_{0}^{\infty} \tau F(\tau ) d\tau .
\end{equation}
In the case of a rectangular potential barrier 
it is given by \cite{fertig}, 
\begin{eqnarray}
\label{reso}
     \left< \tau \right>
     & = & \frac{m}{\hbar} \frac{2k}{\kappa}
       \frac{A \kappa d - B \sin \kappa d \cos \kappa d }
       {B^2 \sin^2 \kappa d + 4k^2 \kappa^2} \nonumber \\
     && \quad 
       + \: i \frac{m}{\hbar}
       \frac{B \kappa d \cos \kappa d
       - A \sin\kappa d}
       {B^2 \sin^2 \kappa d + 4k^2 \kappa^2} \:
       \frac{B}{\kappa^2} \sin\kappa d ,
\end{eqnarray}
where 
\begin{equation}
 A \equiv k^2+\kappa^2, \quad B \equiv k^2-\kappa^2 .
\end{equation}

Under the resonance (the constructive interference) 
conditions $\kappa d=n\pi \: (n=1,2,\cdots )$, 
Equation~(\ref{reso}) is reduced to the expression
\begin{equation}
 \left< \tau \right> = \frac{mdA}{2\hbar k\kappa^2} .
\end{equation}
Using the inequality 
\begin{equation}
 \frac{\alpha + \beta}{2} \geq \sqrt{\alpha \beta} \quad 
 (\forall \alpha , \beta >0) ,
\end{equation}
we can prove that for any condition, $\left< \tau \right>$
is larger than the classical traversal time $\tau_{\rm cl}$
defined by $\tau_{\rm cl}\equiv md / \hbar \kappa$.
That is, $\left< \tau \right> \ge \tau_{\rm cl}$.
As well, the imaginary part of the mean traversal time vanishes
under these conditions.

On the other hand, destructive interference occurs at
$\kappa d = (n+1/2) \pi$. 
Under these conditions, the mean traversal time is 
\begin{equation}
 \left< \tau \right>
  = \frac{2mkd}{\hbar A} - i \frac{mB}{\hbar \kappa^2 A} .
\end{equation}
In particular, we have
\begin{equation}
 {\rm Re} [ \left< \tau \right> ] 
  = \frac{2mkd}{\hbar A} \leq \tau_{\rm cl} .
\end{equation}
Hence, the real part of $\left< \tau \right>$
is smaller than the classical traversal time for any 
case of destructive resonance.

\begin{figure}
     \centering
     \includegraphics[width=0.7\linewidth]{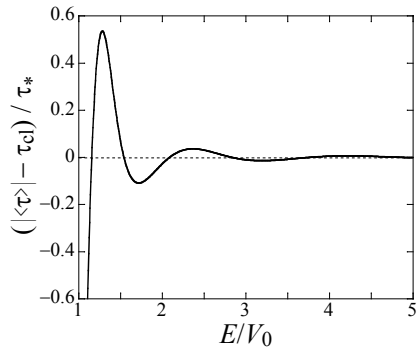}
     \caption{\label{fig:a-tau}
        Deviation of the mean traversal time
        from the classical traversal time,
        $\left| \left< \tau \right> \right| - \tau_{\rm cl}$,
        in the absence of dissipation.
        In the calculation, we adopted $d/\lambda_{0}=5$.
        The resonance points are given by
        $E/V_{0} = 1 + n^2 \pi^2 / (d/\lambda_{0})^2 \simeq 1.39$,
        $2.58$, $4.55$, $\cdots$.}
\end{figure}

Figure~\ref{fig:a-tau} shows
$| \left< \tau \right> | - \tau_{\rm cl}$ as a function of $E/V_{0}$.
There are several peaks around the resonance conditions.
The absolute value $| \left< \tau \right> |$ approaches 
the classical traversal time for larger $E$.
Thus, the particle experiences a longer traversal time 
under resonance conditions as a result of
multiple quantum reflections at the boundaries.
This explains why nonuniform suppressions of 
transmission probabilities in the presence of dissipation occur. 
That is, the longer traversal time 
leads to larger dissipation 
because the damping factor $f(t)$ depends on 
the traversal time through region II.
The most significant reduction in the transmission probability 
is then expected to occur at around 
the first resonance resulting from the longest traversal time compared to 
the classical one  
as shown in Fig.~\ref{fig:a-tau}. 

This can be seen in the formula of the transmission amplitude
$w_{\rm D}(E,V_{0})$ with dissipation Eq.~(\ref{eq:tr-amp}).  
It can be rewritten as
\begin{equation}
\label{eq:w_D-F_tau}
 w_{\rm D} (E,V_{0}) 
 = w (E,V_{0}) \int_{0}^{\infty} d\tau
   \sqrt{f(\tau )} F(\tau ) .
\end{equation}
In the absence of dissipation, we can reproduce 
the non-dissipative expression 
since $\int_{0}^{\infty} d\tau F(\tau ) = 1 $. 
The function $f(\tau )$ is a positive-definite 
decaying function as mentioned above. Hence, 
the factor $\sqrt{f(\tau )}$ in Eq.~(\ref{eq:w_D-F_tau}) 
is considered to weaken the contribution 
from paths with a longer traversal time through region II.  
Therefore, paths with a shorter traversal time
mainly contribute to  the probability amplitude
in the presence of dissipation.

\subsection{Traversal time distribution 
            in the presence of dissipation}
\label{sec:tt-d}

The dissipative effect on the traversal time 
also appears in traversal time distribution. 
Based on the same formulation\cite{fertig}, 
we can evaluate the traversal time distribution in the presence 
of dissipation by replacing $w$ with $w_{\rm D}$. 
The probability amplitude is then given by 
\begin{eqnarray}
\label{eq:f-tau-d}
 F_{\rm D}(\tau ) =  \frac{1}{w_{\rm D}(E,V_{0})}
       \int_{-\infty}^{\infty} \frac{d\omega}{2\pi}
       e^{-i\omega \tau} w_{\rm D}(E,V_{0}-\hbar \omega) ,
\end{eqnarray}
Note that the traversal time distribution is related to  
the potential variation given by the right-hand side 
in Eq.~(\ref{eq:f-tau-d}), which implies an equivalence
between Fig.~\ref{fig:potential}(c) 
and Fig.~\ref{fig:potential}(d). 
After some calculations, we obtain
\begin{equation}
 F_{\rm D}(\tau ) 
     = \frac{\sqrt{f(\tau )} F(\tau )}
        {\int_{0}^{\infty} \sqrt{f(\tau )} F(\tau ) d\tau} .
\end{equation}
 From the above-mentioned properties of $\sqrt{f(\tau)}$, 
the center of the distribution function $F_{\rm D}(\tau ) $ is 
relatively shifted to the direction of smaller $\tau$
when compared with $F(\tau )$ without dissipation.
Namely, the mean traversal time $\left< \tau_{\rm D} \right>
= \int_{0}^{\infty} \tau F_{\rm D}(\tau ) d\tau$ 
is considered to become smaller than $\left< \tau \right> $.
This feature may be {\it counterintuitive} with respect to 
the idea that dissipation causes particle slow down.

\begin{figure}
     \centering
     \includegraphics[width=0.7\linewidth]{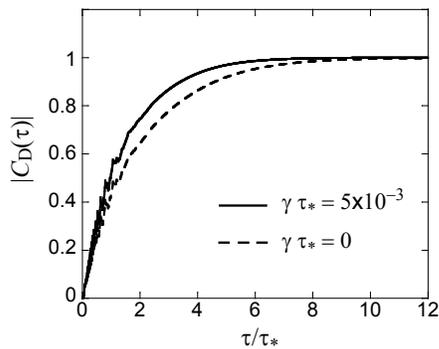}
     \caption{\label{fig:ct}
        The absolute value of the
        cumulative probability amplitude $C_{\rm D}(\tau )$.
        We adopted $E/V_{0} = 1.3$, $d/\lambda_{0}=5$,
        $\Omega \tau_{\ast} = 100$ and
        $\gamma \tau_{\ast} = 5\times 10^{-3}$ or $0$.}
\end{figure}

This feature can be also found in 
the cumulative probability amplitude 
$C_{\rm D}(\tau)$ defined as
\begin{eqnarray}
 \label{eq:ct}
 C_{\rm D}(\tau) & = & \int_0^{\tau}d\tau' F_{\rm D} (\tau') 
    \nonumber \\
 & = & \frac{1}{w_{\rm D}(E,V_{0})}
       \int_{-\infty}^{\infty} d\omega
       \frac{\sin \omega \tau}{\pi \omega} 
  w_{\rm D}(E,V_{0}-\hbar \omega) ,
\end{eqnarray}
which describes the probability amplitude for the traversal time
taking a value between $0$ and $\tau$.
The function $C_{\rm D}(\tau)$ asymptotically approaches 
unity for a larger $\tau$.
Figure \ref{fig:ct} shows examples of
the cumulative probability amplitude $C_{\rm D}(\tau )$ 
for $\gamma \neq 0$ and $\gamma=0$.
Two curves converge to 1 by definition as $\tau$ becomes larger.
In particular, $C_{\rm D}(\tau )$ with a non-zero $\gamma$
converges to 1 faster than the curve for $\gamma=0$.
This means that the traversal-time distribution
becomes narrower and the mean value becomes smaller
when the dissipative effect is taken into account.
This reduced mean value arises from the 
path selection caused by the dissipative effect, whereby
paths taking a longer time in region II
are effectively discarded.
Therefore, the mean traversal time becomes shorter.

\section{Summary and Discussion}
\label{sec:summary}

We have studied the effect of dissipation on the quantum transmission 
of a particle through a rectangular potential barrier, 
especially focusing on transmission resonance. 
We extended the path decomposition method to 
incorporate the dissipative effect into the calculations of
the quantum transmission. 
The transmission probabilities are always suppressed 
by the effect of dissipation, especially at the first resonance 
because of the longer traversal time. 
As well, the mean traversal time in the presence of dissipation
becomes smaller than that in a non-dissipative case. 
This is the result of path selections due to dissipation. 
We have not restricted ourselves to a specific scale. 
Thus our theory is applicable to any scale, including a macroscopic one. 
The study of transmission resonance of a macroscopic object 
is useful to test an applicability of quantum mechanics 
on a macroscopic scale.

Finally, we discuss an experimental setup for such a 
macroscopic object to test our theory. 
Such an experiment could be realized in a specific 
{\it macroscopic} system.
A promising candidate is a system consisting of
a fluxon in a long Josephson junction. 
The fluxon is a topological soliton excitation with a quantum 
unit of magnetic flux produced by a circulating supercurrent, 
i.e., a vortex, and is  
regarded as a single free macroscopic particle 
characterized by a huge number of microscopic degrees
of freedom.\cite{sg-soliton}
It also behaves like a {\em quantum} particle\cite{ki}
in a mesoscopic Josephson junction with 
small capacitance per unit area.
In fact, the quantum tunneling of a fluxon 
has recently been observed in a long annular 
Josephson junction\cite{wallraff}.

Under these circumstances, a fluxon transmission experiment is possible.  
The potential barrier for the fluxon can be made of 
a microshort,\cite{Kivshar}, 
which is a part made of a thinner insulator than the other part. 
The study of fluxon transmission will complement 
that of quantum tunneling 
in research on macroscopic quantum phenomena.
Moreover, a fluxon transmission experiment will also 
provide an important basis for implementing 
quantum computation in superconducting nanocircuits.
Indeed, a qubit using superposition states 
of fluxons or breathers has recently been considered \cite{Fujii}.

\section*{Acknowledgments}

This work was supported in part by a Grant-in-Aid
for Scientific Research from The 21st Century COE
Program ``Topological Science and Technology'',
by a Grant-in-Aid for Scientific 
Research (18540352) from the Ministry of Education Culture, Sports, 
Science and Technology of Japan
and by JSPS KAKENHI (17740267). 
One author (K.K.) thanks
Y. Asano for useful conversations.
The numerical calculations were carried out 
on computers at YITP in Kyoto University.

\appendix

\section{Propagator and its Green's function}
\label{sec:pg}

In the energy representation, 
the propagator is expressed by the Fourier transformation as
\begin{equation}
\label{eq:define-green-f}
 G (x_{T} , x_{0} ; E)
 \equiv i \int_{0}^{\infty} dT K (x_{T},x_{0} ; T)
 e^{iET/ \hbar} .
\end{equation}
Equation~(\ref{eq:path-d}) is now reexpressed in this representation as  
\begin{eqnarray}
\label{eq-g-exp}
 \lefteqn{G(x_{T},x_{0} ; E)} \nonumber \\
 & = & G^{({\rm I})} (a,x_{0} ; E) \Sigma_{x}^{a} 
    \left( G^{({\rm II})} (b,x; E) \right)
  \Sigma_{x}^{b} 
    \left( G^{({\rm III})} (x_{T},x ; E) \right)
   \nonumber \\
 && \quad + G^{({\rm I})} (a,x_{0} ; E) \Sigma_{x}^{a} 
    \left( G^{({\rm II})} (a,x; E) \right)
  \Sigma_{x}^{a} 
    \left( G^{({\rm I})} (a,x ; E) \right)
   \nonumber \\
 && \qquad \times \Sigma_{x}^{a} 
    \left( G^{({\rm II})} (b,x; E) \right)
  \Sigma_{x}^{b} 
    \left( G^{({\rm III})} (x_{T},x ; E) \right)
  + \cdots .
\end{eqnarray}
In the case of the rectangular potential,
the Green functions are given by \cite{fertig}
\begin{eqnarray}
 G^{({\rm I})} (x',x) 
  & = & -\frac{2m}{\hbar k} e^{-ik(x-a)} \sin k (x'-a) , \\
\label{eq:GII}
 G^{({\rm II})} (x',x) 
  & = & -\frac{2m}{\hbar \kappa} 
    \frac{\sin \kappa (x-a) \sin \kappa (x'-b)}
    {\sin \kappa d} , \\
 G^{({\rm III})} (x',x) 
  & = & \frac{2m}{\hbar k} e^{ik(x'-a)} \sin k (x-a) ,
\end{eqnarray}
where $x < x'$, $d\equiv b-a$ and
\begin{equation}
 k \equiv \frac{\sqrt{2mE}}{\hbar}, \quad 
 \kappa \equiv \frac{\sqrt{2m(E-V_{0})}}{\hbar} .
\end{equation}
Here $m$ is the particle mass.
Using these Green functions, Eq.~(\ref{eq-g-exp})
can be calculated in the form \cite{fertig}
\begin{equation}
 G(x_{T},x_{0} ; E) = w(E,V_{0}) G_{0} (x_{T},x_{0}) ,
\end{equation}
where $G_{0} (x_{T},x_{0})$ is the Green function
when the barrier is absent
\begin{equation}
\label{eq:greenf_0}
 G_{0} (x_{T},x_{0}) = - \frac{m}{i \hbar k} e^{ik(x_{T}-x_{0})} ,
\end{equation}
and $w(E,V_{0})$ is the transmission amplitude given by 
\begin{eqnarray}
\label{eq-w}
 w(E,V_{0})
 = - \frac{2i k \kappa e^{-ikd}}
 {\left( k^2 + \kappa^2 \right) \sin \kappa d 
  + 2ik\kappa \cos \kappa d} .
\end{eqnarray}
The usual expression for the transmission probability
is obtained as 
\begin{eqnarray}
 | w(E,V_{0}) |^2
 = \frac{4 k^2 \kappa^2}
 {\left( k^2 - \kappa^2 \right)^2 \sin^2 \kappa d 
  + 4k^2\kappa^2} .
\end{eqnarray}
Thus the exact transmission probability can also 
be obtained using the path decomposition expansion.

\section{Free Particle coupled to Harmonic Oscillators}
\label{sec:dfp}

Caldeira \& Leggett\cite{cl} published a result for  
the quantum Brownian motion in a harmonic potential.
The result is modified in the following way for a free particle,
\begin{eqnarray}
\label{eq:j-final}
 \lefteqn{J(x_{\rm f}, y_{\rm f},t; x_{\rm i}, y_{\rm i},0)} 
  \nonumber \\
 & = & \frac{N(t)}{\pi \hbar} \exp \left[ \frac{i}{\hbar} \left\{
  K(t) \left( \zeta_{\rm f} \xi_{\rm f}
  + \zeta_{\rm i} \xi_{\rm i} \right) - L(t) \zeta_{\rm i} \xi_{\rm f}
  \right. \right. \nonumber \\
 && \quad \left. \left.
  - N(t) \zeta_{\rm f} \xi_{\rm i} - \gamma \frac{m}{2}
  \left( \zeta_{\rm f} \xi_{\rm f}
  - \zeta_{\rm i} \xi_{\rm i} \right) \right\}\right]
  \nonumber \\
 &&  \quad \times \exp \left[ - \frac{1}{\hbar}
  \left\{ A(t) \xi_{\rm f}^{2} 
             + B(t) \xi_{\rm f} \xi_{\rm i} 
             + C(t) \xi_{\rm i}^2  \right\} \right] , 
\end{eqnarray}
where 
\begin{eqnarray}
\label{eq:I_t}
 K(t) & = & \frac{m}{2} \sigma
   \coth \sigma t, \\
\label{eq:N_t}
 L(t) & = & \frac{m}{2} \frac{\sigma
  e^{-\gamma t}}{\sinh \sigma t}, \\
\label{eq:O_t}
 N(t) & = & \frac{m}{2} \frac{\sigma
  e^{\gamma t}}{\sinh \sigma t} , \\
\label{eq:P_t}
 A(t) & = & \frac{2m\gamma}{\pi} \int_{0}^{\Omega} d\omega
  \omega \coth\frac{\hbar \omega}{2k_{\rm B}T'} \int_{0}^{t}d\tau
  \int_{0}^{\tau} ds \nonumber \\ 
 && \quad \times e^{\gamma (\tau + s -2t)}
  \frac{\cos \omega (\tau - s)
  \sinh \sigma \tau 
  \sinh \sigma s}
  {\sinh^2 \sigma t} , \\ 
\label{eq:R_t}
 B(t) & = & - \frac{2m\gamma}{\pi} \int_{0}^{\Omega} d\omega
  \omega \coth\frac{\hbar \omega}{2k_{\rm B}T'} \int_{0}^{t}d\tau
  \int_{0}^{\tau} ds \nonumber \\ 
 && \quad \times  e^{\gamma (\tau + s -t)} 
  \frac{\cos \omega (\tau - s)}
  {\sinh^2 \sigma t} 
  [ \sinh \sigma \tau 
  \sinh \sigma (s-t) \nonumber \\
 && \quad + \sinh \sigma (\tau-t) 
  \sinh \sigma s ] . \\
\label{eq:Q_t}
 C(t) & = & \frac{2m\gamma}{\pi} \int_{0}^{\Omega} d\omega
  \omega \coth\frac{\hbar \omega}{2k_{\rm B}T'} \int_{0}^{t}d\tau
  \int_{0}^{\tau} ds e^{\gamma (\tau + s)}\nonumber \\ 
 && \quad \times \frac{\cos \omega (\tau - s)
  \sinh \sigma (\tau - t ) 
  \sinh \sigma (s-t)}
  {\sinh^2 \sigma t} , 
\end{eqnarray}
where $J(x_{\rm f},y_{\rm f},t;x_{\rm i},y_{\rm i},0)$ is 
the propagator for the density matrix $\rho (x,y,t)$ of 
the free particle, i.e., 
$\rho (x,y,t) = \int dx'dy' J(x,y,t;x',y',0) \rho (x',y',0)$.
Here, $k_{\rm B}$ is the Boltzman constant, 
$T'$ is the temperature, $\zeta \equiv x+y$, $\xi \equiv x-y$, 
and the boundary condition is given by
$\zeta (0) = \zeta_{\rm i}$, $\zeta (t) = \zeta_{\rm f}$ 
$\xi (0) = \xi_{\rm i}$ and $\xi (t) = \xi_{\rm f}$.

\end{document}